\documentstyle[preprint,aps,prb]{revtex} 
\begin{document}
\draft
\preprint{September 19, 1995}


\title{Effect of quantum fluctuations on structural phase transitions in
SrTiO$_3$ and BaTiO$_3$}

\author{W.~Zhong and David Vanderbilt}

\address{Department of Physics and Astronomy
  Rutgers University, Piscataway, NJ 08855-0849}

\date{\today}
\maketitle


\begin{abstract}
Using path-integral Monte Carol simulations and an {\it ab initio}
effective Hamiltonian, we study the effects of quantum fluctuations
on structural phase transitions in the cubic perovskite compounds
SrTiO$_3$ and BaTiO$_3$.  We find quantum fluctuations affect
ferroelectric (FE) transitions more strongly than antiferrodistortive
(AFD) ones, even though the effective mass of a single FE local
mode is larger.  For SrTiO$_3$ we find that the quantum fluctuations
suppress the FE transition completely, and reduce the AFD transition
temperature from 130K to 110K.  For BaTiO$_3$, quantum fluctuations
do not affect the order of the transition, but do reduce the
transition temperature by 35--50 K.  The implications of the
calculations are discussed.
\end{abstract}

\narrowtext

Quantum fluctuations typically have a very important effect on the
structural and thermodynamic properties of materials consisting of
light atoms like hydrogen and helium.  For example, quantum effects
introduce large corrections to the calculated hydrogen
density distribution in the Nb:H system.\cite{piapp}
For materials with heavier atoms, however, the quantum fluctuation can
have only a small effect on the distribution of atomic displacements,
and thus typically do not have a
noticeable effect on the structural and thermodynamic properties of the
material.
However, exceptions may occur.  As we shall see, the cubic perovskites
can exhibit decisive quantum-fluctuation effects, despite the fact
that the lightest constituent is oxygen.  This can occur because
these materials have several competing structures with very small
structural and energetic differences.\cite{lines}

A good example is SrTiO$_3$.  While it has the simple
cubic perovskite structure at high temperature, SrTiO$_3$ goes
through an antiferrodistortive (AFD) transition at 105K to a
tetragonal phase in which the oxygen octahedra have rotated in
opposite senses in neighboring unit cells.
The observed softening of the ferroelectric (FE) polar phonons
with further reduction of temperature in the range 50-100K
would appear to extrapolate to a FE transition close to
20K, but instead the softening saturates and no such transition
is observed.\cite{viana}
The absence of a true FE transition is suggested to be suppressed
by quantum fluctuations, giving rise to a ``quantum paraelectric''
phase at very low temperature.\cite{muller}  Some experiments
appear to suggest a sharp transition to this low-temperature
phase at about 40K, perhaps indicating the formation of some kind
quantum coherent state.\cite{muller2,vacher} However, until a
plausible candidate for the order parameter of the low-temperature
phase is put forward, these ideas must remain highly speculative.

These developments have stimulated
many theoretical efforts to understand the quantum effects in
SrTiO$_3$.\cite{muller,mart,barrett,schneider,martonak}
However, the previous work has all been qualitative or
empirical in approach.  Although it was shown that quantum zero-point
motion is capable of suppressing phase transitions,\cite{schneider}
a detailed microscopic approach is needed to gain a quantitative and
detailed understanding of the quantum effects at finite temperature.
Recently, an {\it ab initio} effective-Hamiltonian scheme has been
developed to study structural phase transitions of cubic perovskites.
It has been successfully applied to BaTiO$_3$\cite{zhong2,zhong2b} and
SrTiO$_3$,\cite{zhong3,zhong3b} giving good agreement with experimental
observations.  Treating atomic motion classically, it predicted FE
phase transitions for SrTiO$_3$ at low temperature, thus giving
indirect support for the notion that quantum fluctuations (not included
in the theory) must be responsible for the observed absence of a
low-temperature FE phase.

In the present work, we have extended the previous treatment of the
first-principles based effective Hamiltonian to include quantum
fluctuations.  In particular,
we use path-integral (PI) quantum Monte Carlo simulations to study
the effect of quantum fluctuations on the structural phase transitions
in SrTiO$_3$ and BaTiO$_3$.  For SrTiO$_3$, we find that the quantum
fluctuations have only a modest effect on the AFD transition temperature,
while the FE transition is suppressed entirely.
We discuss the relative importance of
AFD and FE quantum fluctuations in some detail, and examine the
potential implications of our results for understanding the low-temperature
behavior of the material.   For BaTiO$_3$, in which
the FE transitions occur at higher temperature, we find that the
quantum effects are less dramatic.

We start by reviewing the effective Hamiltonian and its construction.
Two approximations are involved.  First, since both the FE and AFD
transitions involve only small structural distortions, we represent the
energy surface by a Taylor expansion around the high-symmetry cubic
perovskite structure, including up to fourth-order anharmonic terms.
Second, because only low-energy distortions are important to the
structural properties, we include only three such distortions in
our expansion: the soft FE mode, the AFD mode, and an elastic mode.
These are represented, respectively, by local-mode amplitudes
${\bf f}_i$, ${\bf a}_i$, and ${\bf u}_i$, where $i$ is a cell index.
The local modes that are constructed in such a way that a uniform
(or, for AFD, a uniformly staggered) arrangement of the mode vectors
represents the desired low-energy excitation.\cite{zhong2b}
Thus, we work with local-mode vectors instead of atomic displacements.
This reduces the number of degrees of freedom from 15 to 9 per cell and
greatly reduces the complexity of the Taylor expansion.
The Hamiltonian is specified by a set of
expansion parameters determined using highly accurate first-principles
calculations with Vanderbilt ultrasoft pseudopotentials.\cite{vand1}
The details of the Hamiltonian, the first-principles calculations, and
the values of the expansion parameters have been reported elsewhere.
\cite{zhong2,zhong2b,zhong3,zhong3b}

In our previous work, we have used this effective Hamiltonian by
applying Monte Carlo simulation techniques to study the thermodynamics
of the system in the classical limit.  Assuming the ionic motions
are classical is usually a good approximation for systems such as
cubic perovskites containing atoms no less massive than oxygen.
However, the structural differences and energy barriers between
the cubic structure and the possible (rhombohedral or tetragonal)
distorted structures are very small.  A rough estimate of the importance
of quantum fluctuations can be obtained from the Heisenberg
uncertainty principle $\Delta p \cdot \Delta q \ge \hbar /2 $,
or equivalently,
\begin{equation}
 \Delta E \ge \hbar^2/(8m\Delta q^2) \; .
\label{eq1}
\end{equation}
Here, $\Delta q$ denotes the uncertainty in the structural coordinate,
which is related to the
structural difference between phases.  $\Delta E$ is the
energy uncertainty, or zero-point energy, which may prevent
the occurrence of the distorted
phase if it is larger than the classical free energy reduction.
So if the structural and energetic differences between phases
are small enough, quantum suppression may occur even for fairly
massive ions.  For a quantitative understanding, we need to perform
statistical simulations that treat the ionic motion quantum-mechanically.

Here, we adopt the path-integral (PI) technique\cite{pi} of quantum
simulations, which has proven to be a very successful method for
studying H- and He-related systems.\cite{piapp,ceperley}
The method is based on Feynman's PI formulation of
quantum mechanics.\cite{feynman}  This formulation states that the
partition function of the original quantum-statistical systems of particles
can be approximated by the partition function of $P$ subsystems of
classical particles with each quantum particle replaced by a cyclic
chain of $P$ beads coupled by harmonic springs.  Each subsystem
(comprising one bead from each chain) has internal interactions identical
to the reference classical system, except for a reduction in strength
by a factor $1/P$.  The spring
constant of the harmonic springs coupling the beads inside a certain
cyclic chain is $mP/\hbar^2\beta^2$, where $m$ is the mass of the
quantum particle and $\beta$ the inverse temperature $(k_BT)^{-1}$.
This approximation becomes exact when the number of beads $P
\rightarrow \infty $, but in practice almost exact results can be
obtained with a finite $P$ depending on the system of interest.
This way,
thermodynamic properties of the $N$-particle quantum system can be
obtained from the study of a $(P \times N)$-particle classical
system.

The only extra inputs we need are the masses of all the ``particles'' in
our system.  The degrees of freedom in our Hamiltonian are the three
local-mode amplitude vectors ${\bf f}_i$, ${\bf a}_i$, and ${\bf u}_i$
associated with each unit cell $i$.  Each local mode involves
displacements of several ions.  If we regard each local vector as
representing the displacement of some ``pseudo-particle,''
the mass of each such pseudo-particle can be determined from all
the ionic displacements involved.  Since two local-mode vectors may
involve the same
ion, we actually have a non-diagonal mass matrix.  For example, the
mass matrix elements between local modes ${\bf f}_i$ and ${\bf f}_j$,
or equivalently, $f_{i\alpha}$ and $f_{j\beta}$, can be constructed
through
\begin{equation}
  m_{i\alpha,j\beta} = \xi (i\alpha) \cdot M \cdot \xi (j\beta) \; .
\end{equation}
Here, $i$ and $j$ are the cell indices, while $\alpha$ and $\beta$
denote Cartesian components.  $\xi (i\alpha)$ is the eigenvector
describing atomic displacements associated with local mode $f_{i\alpha}$,
and $M$ is a (diagonal) mass matrix in the 15$L^3$-dimensional space of
atomic displacements of our $L\times L\times L$ supercell.
Similarly, mass matrix elements connecting different kinds of local
vectors, such as those between ${\bf f}_i$ and ${\bf a}_i$, are also
included.  The entire mass matrix can be calculated once and for all,
and the extension of the PI technique to handle a non-diagonal mass
matrix is straightforward.

The study of the thermodynamic properties of the classical system is
performed using Monte Carlo (MC) simulations.\cite{MC} The original
simulation cell is an $L \times L \times L$ cube, with three vectors
${\bf f}_i$, ${\bf a}_i$, and ${\bf u}_i$ at each lattice point $i$.
Periodic boundary conditions are used, and homogeneous strains of the
entire supercell are included.  Each local vector is converted to a
string of $P$ beads, so that we have $9PL^3$ degrees of freedom per
simulation supercell.  We use a single-flip algorithm, making trial
moves of the vectors at each site in turn and testing acceptance
after each move.  We say that one Monte Carlo sweep (MCS) has been
completed when all vectors on all sites have been tried once.
Because of the 1\% lattice-constant error in our LDA calculations
and the strong sensitivity of the structural transitions to the
lattice constant, all our simulations are performed at a negative
pressure to restore the experiment lattice constant, as in our
previous work.\cite{zhong2,zhong2b,zhong3,zhong3b}

The Trotter number $P$ should be large enough to ensure that the
quantum effects are correctly accounted for.  On the other hand, the
computational load increases rapidly with increasing $P$, because of
both larger system size and longer correlation time with larger $P$.
In our simulation, the proper Trotter number for each temperature is
chosen empirically.  For a certain temperature, we simulate systems
with increasing Trotter number $P=1,2,4,8,16, ...,$. We equilibrate
systems with each $P$ and monitor their order parameters.  We determine
that the $P$ is large enough if the monitored quantities converge.  If
a certain quantity is sensitive to $P$,  its value at $P=\infty$ can be
extrapolated following the formula $a_0+a_1/P+a_2/P^2$ (Ref.
\onlinecite{cuccoli}).

We concentrate on SrTiO$_3$ and study the effect of quantum fluctuation
on both FE and AFD phase transitions.  In Fig. \ref{u_T}, we show the FE
and AFD order parameters {\bf f}($\Gamma$) and {\bf a}($R$) as a
function of temperature for a $12 \times 12 \times 12$ simulation
cell.  The classical data (previously published in
Ref.\onlinecite{zhong3}) are produced by a cooling-down simulation,
starting at 250K and cooling down gradually, equilibrating and then
simulating to obtain the order parameters.\cite{zhong3}  The quantum
simulations are performed with $P=4$, which is found to give converged
results for $T>60$K and qualitatively correct results for $T>20$K.  We
use the equilibrium configuration from the classical simulations
($P$=1) as the starting configuration.  We find
the system reaches equilibrium faster this way than it does if
gradually cooled and the results are less affected by hysteresis.  The
system is equilibrated for 10,000 MCS's, and then another 30,000
to 70,000 MCS's are used to obtain the reported thermodynamic averages.

Fig.\ \ref{u_T} shows that the quantum fluctuations do affect both
the AFD and FE phase transitions.  The AFD phase transition
temperature decreases from 130K to 110K when the quantum
fluctuations are turned on, bringing the results into better
agreement with the experimental result of 105K.  On the other hand,
the quantum fluctuations can be seen to have completely suppressed
the FE phase transitions, at least down to 40K.  Further
simulations going as high as $P=20$ place an upper bound of about
5K on any possible FE phase transition temperature.  Thus, we
conclude that quantum fluctuations almost certainly suppress the FE
phase transitions completely, resulting in a paraelectric phase
down to $T=0$.

Since the effect of quantum fluctuations is more dramatic on the
FE transitions, we analyze this case in more detail.  In the
paraelectric phase, the fluctuation of the FE local-mode vector
{\bf f} has both quantum and thermal contributions.  We identify
the thermal fluctuations as those associated with the fluctuations
of the center of gravity of the cyclic chain. More specifically,
letting ${\bf f}(i,s,t)$ represent $\bf f$ on lattice site $i$,
Trotter slice $s$, and MCS $t$, the thermal fluctuation can be
obtained from our simulation using
\begin{equation}
(\Delta f^{\rm thermal})^2 = \left\langle \left\langle
f \right\rangle_{s}^2 \right\rangle_{i,t} \; ,
\end{equation}
while the total fluctuation is
\begin{equation}
(\Delta f^{\rm total})^2 = \left\langle f^2 \right\rangle_{i,s,t} \; .
\end{equation}
Here the bracket represents the indicated average.  The part of fluctuation
due solely to the quantum effects can be obtained from $ (\Delta f^{\rm
QM})^2 = (\Delta f^{\rm total})^2 - (\Delta f^{\rm thermal})^2 $.
The result for a $ 10 \times 10 \times 10 $ lattice is shown in
Fig.\ \ref{f2}.  The results are obtained from simulations at
several small Trotter numbers and then extrapolated to $P=\infty$
using the formula $a_0+a_1/P+a_2/P^2$.  As expected, the thermal
fluctuation decrease with decreasing temperature, while the quantum
fluctuations increase.  Below 70K, the quantum fluctuation
dominate.

Recent experiments suggest there may be a weak signature of a phase
transition in SrTiO$_3$ around 40K.\cite{muller2}  This was tentatively
suggested to be a phase transition to a coherent quantum state in which
small FE domains propagate through the crystal.  Because
the size of our simulation cell is much smaller than the domain size
suggested, we expect that such a state would appear as a real FE phase
in our simulation.  This is not observed.  However, our simulation
does reveal some changes in the character of the FE fluctuations
at low temperature.
A typical FE fluctuation at high temperature resembles the soft
eigenvector of the force-constant matrix, which is independent of
the masses since the classical thermodynamic properties are related
only to the potential energy.  However, the quantum fluctuations
are quite sensitive to the ionic masses, and at low temperature the
fluctuations of light (primarily oxygen-related) degrees of freedom
are accentuated.  This cross-over in the character of the
fluctuations occurs gradually below 100K, and we suspect that it
might possibly be responsible for the experimentally observed
anomalies which were interpreted in terms of a phase transition.
If this is the case, the ``quantum paraelectric'' phase at very low
temperature is probably not separated by a true phase transition
from the classical paraelectric phase at higher temperature.

To better characterize the impact of the quantum effects on FE
transitions, we also apply the PI simulations to BaTiO$_3$.  The
results are summarized in Table \ref{t1}.  The simulation procedure
is the same as for SrTiO$_3$, except that the AFD degrees
of freedom are neglected in BaTiO$_3$ because of their high energy.
Experimentally, BaTiO$_3$ has four phases in the sequence cubic (C),
tetragonal (T), orthorhombic (O), and rhombohedral (R) with decreasing
temperature.  Our classical simulations correctly reproduce this
transition sequence, and give transition temperatures that are
in reasonable agreement with ($\sim$ 15-30\% below) the
experimental ones.  We have argued previously that the quantitative
discrepancy can probably be traced to the LDA lattice-constant
error.\cite{zhong2,zhong2b} Here, we find
that, with quantum effects included, the calculated transition
sequence is still the same, while the transition temperatures are
reduced further by 35K to 50K.  Although the absolute transition
temperatures are thus in slightly worse agreement with experiment,
the spacing between phases is more reasonable.  In any case, it
is clear that the quantum effects can have a substantial effect
on the FE transition temperatures even up to several hundreds of
degrees K, a result which was not obvious from the outset.

It may appear counterintuitive that quantum effects on the FE
instability are much stronger than on the AFD instability in
SrTiO$_3$.  After all, the AFD instability involves only the motion
of oxygen atoms, while the FE instability involves mainly Ti
atoms which are three times heavier than the oxygen atoms.  A
partial explanation can be drawn from the fact that the structural
change involved in the FE distortion (0.1 a.u for Ti in SrTiO$_3$)
is much smaller than for the AFD distortion (0.3 a.u. for O).  As a
result, $m\Delta q^2$ turns out to be three times larger for the
AFD case, even though the effective mass is smaller.  Thus,
according to Eq. (\ref{eq1}), the effect of the quantum
fluctuations will be less significant for the AFD case.

We think a more fundamental explanation may be found in the stronger
spatial correlations between AFD distortions.  In the cubic phase,
the spatial correlations for the FE local vectors are chain-like or
quasi-1D: $f_z ({\bf R}_i) $ correlates strongly only with
$f_z ({\bf R}_i \pm n a \hat{\bf z})$, where $n$ is a small integer
number and $a$ is the lattice constant.\cite{yu,zhong3b} This
correlation is due to the strong Coulomb interactions between FE
local modes,\cite{zhong1} which strongly suppress longitudinal
excitations relative to transverse ones.  With the correlation length
estimated at $10a$,\cite{zhong3b} we can roughly say that about
10 local-mode vectors are ``bound together'' and the effective mass
becomes 10 times larger.  On the other hand, the AFD modes,
associated with rotation of the oxygen octahedral, correlate strongly
with each other because of the rigidity of the octahedral unit.
The correlation region is 2D disc-like:
$a_z ({\bf R}_i)$ correlates strongly with $a_z ({\bf
R}_i \pm n a \hat{\bf x} + m a \hat{\bf y})$, where $m$ is again a
small integer.  The AFD correlation length is comparable with
the FE one,\cite{zhong3b} but now the 2D nature implies that roughly
100 mode-vectors are tied together, for a mass enhancement of 100
instead of just 10.  Thus, this effect weakens the quantum fluctuations
much more for the AFD than for the FE case, and one should
generally expect quantum suppression of phase transitions to be
stronger in the FE case.

In summary, we have applied the PI technique to study the effect of
quantum fluctuations on FE and AFD phase transitions in SrTiO$_3$
and BaTiO$_3$.  We find that the quantum fluctuations have a weaker
effect on the AFD transition than on the FE one, because the AFD
modes are more strongly correlated with each other.  In the case of
SrTiO$_3$, we find that the FE phase is suppressed entirely,
thereby supporting the notion of ``quantum paraelectric'' behavior
(though not necessarily a distinct phase) at very low temperature.
The AFD transition temperature is found to be only slightly
reduced.  For BaTiO$_3$, we find that the quantum effects preserve
the transition sequence and reduce the transition temperatures
modestly.

This work was supported by ONR grant N00014-91-J-1184.


\begin{figure}
\caption{AFD and FE order parameters {\bf a}($R$) and {\bf
f}($\Gamma$) as a function of temperature for a $12 \times 12
\times 12$ SrTiO$_3$ simulation cell.  Squares, circles, and triangles
indicate the largest, intermediate, and smallest components of the
order parameter, respectively.  Filled symbols are from classical
simulations, while open symbols are from path-integral simulations
with $P=4$ (the latter for the FE case are nearly zero and are thus
not very visible).  Insets indicate schematically the nature of the
AFD and FE distortions.
\label{u_T}}
\end{figure}

\begin{figure}
\caption{Classical (squares), quantum (circles), and total (triangles)
RMS fluctuation of the FE local-mode vectors
[Eqs.\ (3-4)] in SrTiO$_3$ as a function of temperature. \label{f2}}
\end{figure}

\begin{table}
\caption{The effect of quantum fluctuations on the FE transition
temperatures in BaTiO$_3$, for a $12\times 12 \times 12$ supercell.
$R$, $O$, $T$, and $C$ indicate rhombohedral,  orthorhombic,
tetragonal, and cubic phases, respectively.
\label{t1}}
\begin{tabular}{clll}
  Phase       &  Classical  &  Quantum  &  Expt.  \\ \hline
  $O-R$       &  200$\pm$ 10 &  150 $\pm$ 10 & 183  \\
  $T-O$       &  232$\pm$ 2   &  195 $\pm$ 5  & 278  \\
  $C-T$       &  296$\pm$ 1   &  265 $\pm$ 5  & 403
\end{tabular}
\end{table}

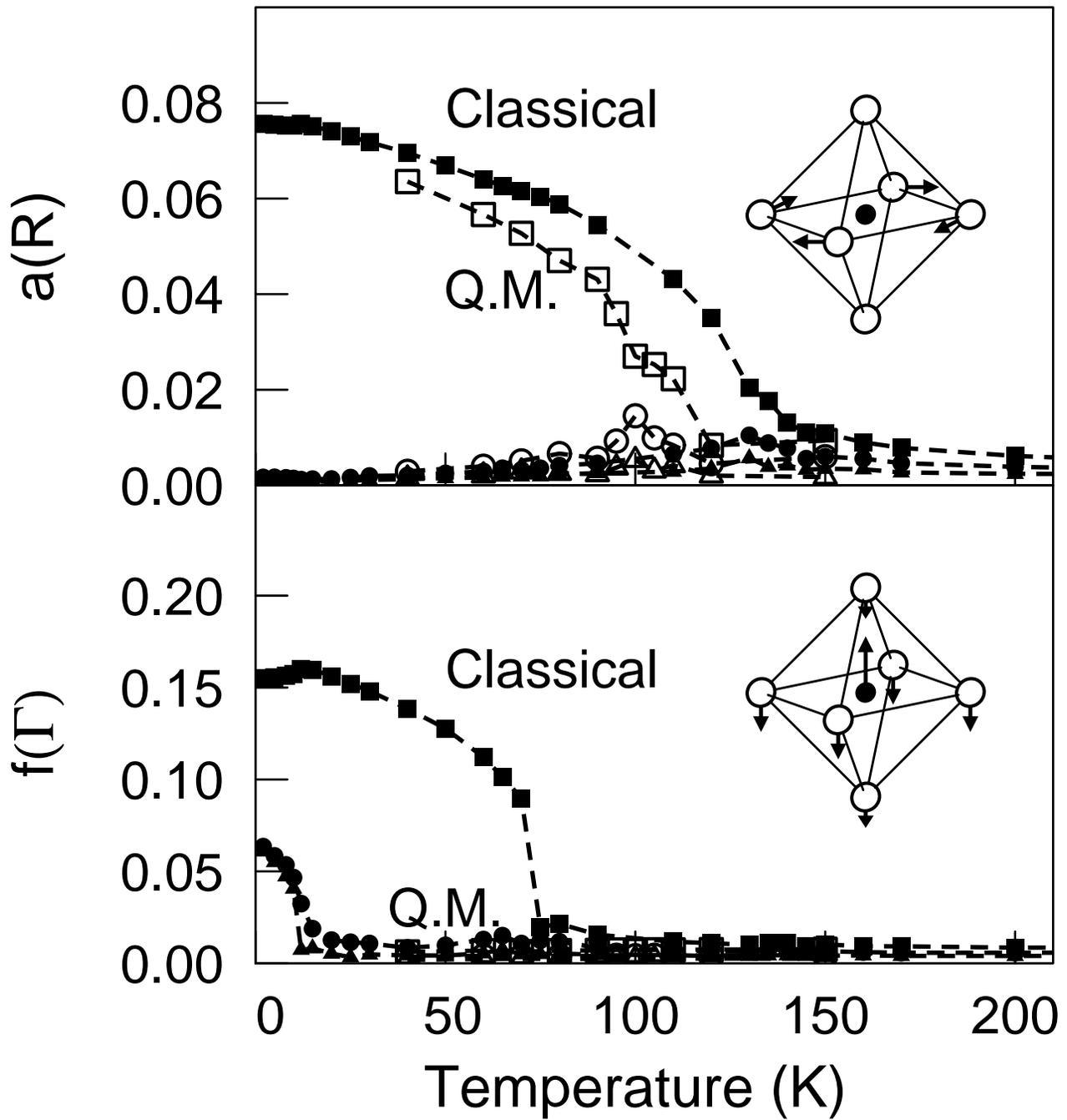
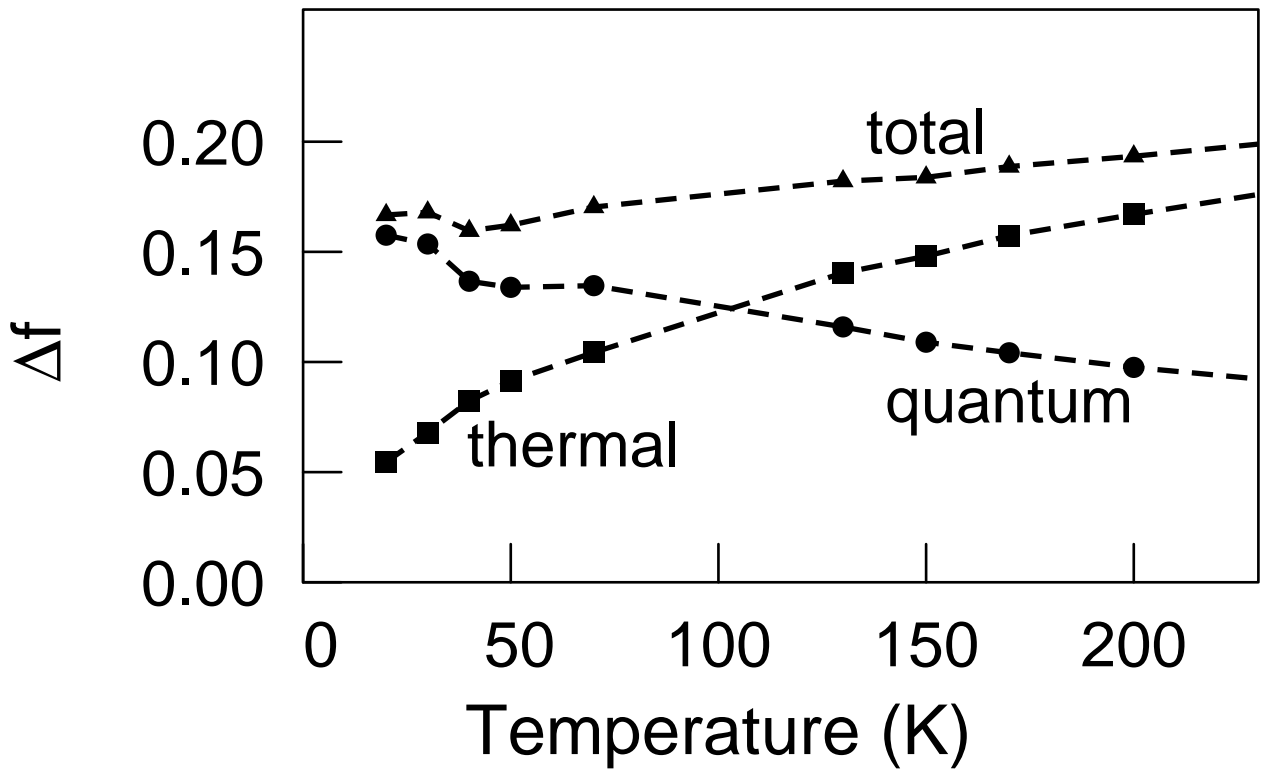
\end{document}